\documentclass[11pt]{article}

\def\thefootnote{\fnsymbol{footnote}}
\newcommand{\myref}[1]{(\ref{#1})}
\newcommand{\myvev}[1]{{\langle #1 \rangle}}
\newcommand{\Gaa}{\Gamma_{\alpha}}
\def\R{\bar R}
\def\del{\delta}
\def\bfr{\mathbf{r}}
\def\bl{\bar{\lambda}}

\def\Del{\Delta}
\def\half{{1\over2}}
\def\bea{\begin{eqnarray}}
\def\eea{\end{eqnarray}}
\def\beq{\begin{equation}}
\def\eeq{\end{equation}}

\def\ux{$U(1)_X$}
\def\uk{$U(1)_K$}
\def\uo{$U(1)$}

\def\vx{V_X}
\def\superint{\int d^{4}\theta}
\newcommand{\WaWa}{ W_a^{\alpha} W^a_{\alpha}}
\newcommand{\Da}{{\cal D}_{\alpha}}

\newcommand{\Dc}{{\cal D}^{\alpha}}
\newcommand{\Dd}{{\cal D}_{\dot{\beta}}}
\newcommand{\Wa}{W_{\alpha}}

\newcommand{\Wc}{W^{\alpha}}

\newcommand{\Xa}{X_{\alpha}}
\newcommand{\Xc}{X^{\alpha}}
\def\re{{\rm Re}}
\def\im{{\rm Im}}

\def\D{{\cal D}}
\def\bD{\bar{\D}}
\def\pp{\partial}

\def\[{\left [}
\def\]{\right ]}
\def\({\left (}
\def\){\right )}
\def\lbr{\left\{}
\def\rbr{\right\}}
\def\r{\right|}
\def\l{\left.}

\def\Z{{\bar{Z}}}
\def\T{\bar{T}}
\def\z{\bar{z}}
\def\S{{\bar{S}}}
\def\Tr{{\rm Tr}}

\def\L{{\cal L}}
\def\s{\bar{s}}
\def\cM{{\cal{M}}}
\def\bF{\bar{F}}
\def\chiproj{(\bD^2 - 8R)}
\def\bchiproj{(\D^2 - 8 \bar R)}

\def\ddd{\nonumber \\ &&}

\def\nnn{\nonumber \\ }
\def\hc{ + {\rm h.c.}}

\begin{document}

\begin{titlepage}
\begin{center}

\hfill UCB-PTH-09/26 \\
\hfill August, 2009 \\[.3in]

{\large{\bf T-DUALITY AND THE WEAKLY COUPLED HETEROTIC 
STRING}}\footnote{Plenary talk at SUSY09, June 10-25, 2009, Northeastern 
University; to be published in the proceedings.}
\footnote{This work was supported in part
by the Director, Office of Science, Office of High Energy and Nuclear
Physics, Division of High Energy Physics, of the U.S. Department of
Energy under Contract DE-AC02-05CH11231, in part by the National
Science Foundation under grants PHY-0457315 and PHY99-07949.}  \\[.2in]

Mary K. Gaillard\\[.1in]

{\em Department of Physics and Theoretical Physics Group,
 Lawrence Berkeley Laboratory, 
 University of California, Berkeley, California 94720}\\[.5in] 

\end{center}

\begin{abstract}
  T-duality is a symmetry of the heterotic string to all orders in
  string perturbation theory.  This results in an effective four
  dimensional supergravity theory with desirable features for
  phenomenology.  T-duality, as well as, generically, an anomalous
  \uo, is broken by quantum anomalies of the effective field theory.
  The structure of the full anomaly is presented, and the mechanisms
  for anomaly cancellation are described.
\end{abstract}
\end{titlepage}

\newpage

\renewcommand{\thepage}{\roman{page}}
\setcounter{page}{2}
\mbox{ }

\vskip 1in

\begin{center}
{\bf Disclaimer}
\end{center}

\vskip .2in

\begin{scriptsize}
\begin{quotation}
This document was prepared as an account of work sponsored by the United
States Government. While this document is believed to contain correct 
 information, neither the United States Government nor any agency
thereof, nor The Regents of the University of California, nor any of their
employees, makes any warranty, express or implied, or assumes any legal
liability or responsibility for the accuracy, completeness, or usefulness
of any information, apparatus, product, or process disclosed, or represents
that its use would not infringe privately owned rights.  Reference herein
to any specific commercial products process, or service by its trade name,
trademark, manufacturer, or otherwise, does not necessarily constitute or
imply its endorsement, recommendation, or favoring by the United States
Government or any agency thereof, or The Regents of the University of
California.  The views and opinions of authors expressed herein do not
necessarily state or reflect those of the United States Government or any
agency thereof, or The Regents of the University of California.
\end{quotation}
\end{scriptsize}

\vskip 2in

\begin{center}
\begin{small}
{\it Lawrence Berkeley Laboratory is an equal opportunity employer.}
\end{small}
\end{center}

\newpage
\renewcommand{\theequation}{\arabic{section}.\arabic{equation}}
\renewcommand{\thepage}{\arabic{page}}
\setcounter{page}{1}
\def\thefootnote{\arabic{footnote}}
\setcounter{footnote}{0}

\section{Introduction}

When compactified from ten to four space-time dimensions, the weakly
coupled heterotic (WCHS) string theory~\cite{wchs} has an invariance
under a discrete group of transformations known as ``T-duality'' or
``target space modular invariance''~\cite{mod}.  This leads to several
attractive features for phenomenology:
\begin{itemize}
\item The K\"ahler moduli, or T-moduli, 
are generically stabilized~\cite{gnrev} at self-dual points:
$t_{\rm s d}\to t_{\rm s d}$; as a consequence there is 
no large flavor mixing induced by supersymmetry (SUSY) breaking.
\item R-symmetry is protected~\cite{bgax} by T-duality in supergravity (SUGRA),
thereby suppressing the mass of the axion. This provides~\cite{gk} a
possible solution to the strong CP problem.
\item When combined with \uo\, gauge symmetries, T-duality
provides~\cite{rparity} a possible mechanism for R-parity or an even
stronger discrete symmetry.
\end{itemize}
I will briefly describe each of these results, which are not new, but serve
as motivation for the second part of this talk, namely anomalies and anomaly
cancellation in SUGRA.

At the quantum level of the effective supergravity theory, T-duality
is broken by quantum anomalies, as is, generically, an Abelian \ux\,
gauge symmetry, both of which are exact symmetries of string
perturbation theory.  It was realized some time ago that these
symmetries could be restored by a combination of four dimensional
counterparts~\cite{gs4} of the Green-Schwarz (GS) mechanism in 10
dimensions~\cite{gs} and string threshold corrections~\cite{th}.
However anomaly cancellation has been demonstrated explicitly only for
the coefficient of the Yang-Mills superfield strength bilinear.  The
entire supergravity chiral anomaly has in fact been
determined~\cite{danf}, but the complete superfield form of the
anomaly is required to fully implement anomaly cancellation.

Chiral anomalies are ill-defined in the unregulated effective
field theory; I use Pauli Villars (PV) regulation~\cite{pv} to define the theory.
Requiring GS anomaly cancellation restricts the form of the anomaly~\cite{bgan},
which in turn leads to constraints on soft SUSY-breaking sfermion masses.
%

\section{The benefits of T-duality}
\subsection{T-moduli stabilization}

T-moduli are generically stabilized at self-dual points: $t_{\rm s d}
= 1,\;e^{i\pi/6}$. To see this\footnote{See Section 3.1.2
of~\cite{gnrev}.}  consider a toy model with a single T-modulus
superfield $T$ and the dilaton superfield $S$.  T-duality and the
shift-symmetry of the axion $\im \l S\r$ require that the K\"ahler
potential $K$ take the form
\beq K = k(S + \S) - 3\ln(T + \T).\eeq
A simple T-duality invariant ansatz for the superpotential $W$, which was
often used in the past, is 
\beq W(S,T) = H(S)\eta^{-6}(T),\eeq
where $\eta$ is the Dedekind eta function. 
Minimization of the potential gives two solutions for the vacuum configuration:  
\bea \langle H_s + k_s H\rangle &=& \langle F_s\rangle = 0,
\qquad H_s = {\pp H\over\pp s}, \quad {\rm etc.,}\label{sol1}\\
\langle(s + \s)^2k_{s s}\rangle &=& \langle - e^{2i\gamma}H^*\[1 +
24\re t\(\re\zeta(t) + 2\re t|\zeta(t)|^2\)\]\rangle,\quad \zeta =
{\pp\ln\eta\over\pp t},\label{sol2}\eea
where $t = \l T\r,\; s = \l S\r$, and $\gamma = \arg(H_s - k_s H)$.
For solution (\ref{sol1}), the self-dual point $t = t_{\rm s d}$ is a local maximum
with $\langle F_s\rangle = 0$; this
is the only  solution in classical limit with $k(S + \S) =
-\ln(S + \S)$.  Solution (\ref{sol2}) instead satisfies $\langle F_t\rangle = 0,$
and $t_{\rm s d}$ is a local minimum. 
For fixed $\langle\re s\rangle$ we can parametrize the dilaton contribution
to the potential as
\beq K^{-1}_{s\s}F^s \bF^{\s} = |2\re s H_s - H|^2 \equiv a|H|^2,\eeq
with $a = 0$ for solution (\ref{sol1}). The potential
\beq V = {H^2\over16\re s(\re t)^3|\eta(t)|^{12}}
\[a + 24\re t\(2\re t|\zeta(t)|^2 + \re\zeta(t)\)\]\eeq 
is was studied in the $\re t$ direction in~\cite{gnrev}. It has the oft-cited
minimum at $t\approx 1.23$ in the classical limit (\ref{sol1}) with
$a=0$, but for $a> .05$, the minimum is always at the self dual point
$t = 1,\;4\re t\zeta(2) = - 1,\; \myvev{F_t}=0$.  This is the result
found~\cite{bgw} in more realistic ``K\"ahler stabilization'' models
for the dilaton.

\subsection{Is the universal string axion the QCD axion?}

If SUSY is broken by a single gaugino condensate
$\myvev{\bl\lambda}\ne0$, there is a residual R-symmetry, and the
axion remains massless at the SUSY breaking scale in the quantum field
theory (QFT) approximation, but in a general SUGRA theory, couplings
of the axion to higher powers $\myvev{(\bl\lambda)^p}$ of the gaugino
condensate may generate a mass that is too large for it to be
identified with the Peccei-Quinn axion~\cite{bd}.

T-duality forbids~\cite{bgax} low values of the exponent $p$. The
minimal group of T-duality transformations, namely $SL(2,{\bf Z})$,
requires $p\ge4$, while, for example, the maximal group for a model
with just three untwisted K\"ahler moduli, namely $[SL(2,{\bf Z})]^3$,
requires $p\ge12$.  An analysis~\cite{gk} of the QCD phase transition
shows that the identification of the string axion $\im s$ with the QCD
axion is possible provided the T-duality group is larger than the
minimal one, requiring $p>4$, with the caveat that it also requires a large
axion coupling parameter, $f_a\sim m_{\rm Pl}$, which may be a viable
possibility~\cite{fpt}. Thus a string solution to the strong CP
problem implies a mild constraint on the group of T-duality
transformations.

\subsection{R-parity?}

If the T-moduli are stabilized at self-dual points, $\myvev{t} 
= {t_{s d}} =1$ or $e^{i\pi/6}$, there is an unbroken discrete subgroup:
\beq G_R = \mathbf{Z}_4^m\otimes \mathbf{Z}_{6}^{m'},\qquad m + m' = 3,\eeq
under which the gauginos $\lambda$ and gauge charged chiral superfields
transform as 
\beq \lambda\to -\lambda,\qquad \Phi^i(\theta)\to 
e^{2\pi i\beta_i(q^i_n)}\Phi^i(\theta'),\label{rtransf}\eeq
where $q^i_n$ is a modular (T-duality) weight. In the presence of an
anomalous \ux\, the corresponding GS-term generates a D-term, resulting
in the breaking of some number $m$ of \uo\, gauge symmetries and of
$G_R$, but leaving an unbroken discrete subgroup $G'_R$
\beq G_R\otimes U(1)^m\to G'_R\in G_R\otimes U(1)^m,\eeq
with the transformation property \myref{rtransf} for chiral superfields
modified by phase factors that depend on their \uo\, charges $q^i_a$:
\beq {\Phi^i(\theta)\to e^{2\pi i\beta_i(q^i_n,q^i_a)}\Phi^i(\theta')}.
\label{rtransf2}\eeq
Finally, at the electroweak gauge symmetry breaking scale:
\beq SU(2)_L\otimes U(1)_{\rm w}\to U(1)_{\rm e m},\qquad \myvev{H_{u,d}} \ne 0,\eeq
the surviving discrete symmetry is the subgroup $R$ that leaves the Higgs fields
invariant:
\beq  R\in G'_R\otimes U(1)_{\rm w},\eeq
with \myref{rtransf2} now replaced by
\beq {\Phi^i(\theta)\to e^{2\pi i\beta_i(q^i_n,q^i_a,Y^i)}\Phi^i(\theta')},
\qquad \beta_{H_{u,d}}=n,\eeq
where $Y^i$ is weak hypercharge.
Requiring nonvanishing quark masses and CKM angles imposes the conditions
\beq \beta_Q = -\beta_{Q^c} \equiv\beta,\eeq
and imposing nonvanishing lepton masses gives 
\beq \beta_L = -\beta_{E^c}\equiv \gamma.\eeq
There will be no dimension-three operators of the type
$$ U^c D'^c D''^c,\quad L Q D'^c,\quad L L' E''^c,$$
provided 
\beq \beta\ne\displaystyle{n\over3},\qquad \gamma\ne n,\eeq
and, in contrast to standard R-parity, the dimension-four operator
\beq U^c U'^c D''^c E^c \eeq
will be forbidden provided
\beq 3\beta + \gamma\ne n.\eeq
A challenge for string model builders is to find a heterotic string vacuum
with the correct modular weights and \uo\, charges to satisfy these conditions.

\section{Anomalies and anomaly cancellation} 

\subsection{Preliminaries}

In conventional superspace, the kinetic Lagrangian for SUGRA and matter superfields
takes the form
\beq \L_{\rm kin} = -3\superint\,E_0 \, e^{-\frac{1}{3}K(Z,\bar{Z})},\label{l0}\eeq
where $E_0$ is the super-determinant of the super-vielbein $E_M^{\, A}$.
By an appropriate K\"ahler and super-Weyl transformation, this
may be put in the form~\cite{bgg}
\beq \L_{\rm kin} = -3\superint\,E\, ,\label{lu1k}\eeq
giving a canonical Einstein term for the component form of the
Lagrangian; this is the K\"ahler~\uo\, [\uk] superspace formulation of
SUGRA. The structure group of \uk\, geometry contains the Lorentz,
$U(1)_K$, Yang-Mills (YM) and chiral superfield reparameterization
groups. A chiral superfield $Z$ is {\it covariantly} chiral: $\Dd Z =
0$, where the covariant spinorial derivative $\Dd$ includes the \uk,
YM, spin and chiral superfield reparameterization connections.  The
K\"ahler potential $K = \Z e^V Z+\ldots$ of the standard formulation
\myref{l0} is replaced simply by $K = |Z|^2+\ldots$ in the \uk\,
superspace formulation \myref{lu1k}. The full Lagrangian takes the form
\bea\L &=& \half\sum_i\superint{E\over R}\bfr_i \hc, \qquad\bfr_{\rm kin} = -3R,\nnn
\bfr_{\rm YM} &=& {1\over4}f(Z)\WaWa,\qquad
\bfr_{\rm superpot} = e^{K/2}W(Z),\label{ltot}\eea
where the superfield
$R$ is a component of the super-Riemann tensor, whose lowest component
$\l R\r$ is an auxiliary field of the SUGRA multiplet; its equation of motion
reads
\beq \l R\r = \half e^{K/2} W(z),\qquad z = \l Z\r,\eeq
and in the WCHS the gauge kinetic function is just the dilaton
superfield: 
\beq f(Z) = S.\eeq  
Local supersymmetry of the Lagrangian \myref{ltot} is
assured~\cite{bgg} by the fact that the superfields $\bfr_i$ have
\uk\, charge $w_K(\bfr) = 2$.

\subsection{PV regularization}

A renormalizable supersymmetric theory is  defined by specifying the 
matter and Yang-Mills chiral superfields $Z^i$ and $\Wa^a$, respectively, 
their gauge transformation properties  
\beq \del^a Z^i = i(T^a Z)^i, \qquad\del^a\Wa^b = f^{a b c}W_{\alpha c},\eeq
and the superpotential $W(Z)$.  The (one loop) ultraviolet (UV)
divergences of the theory can be regulated~\cite{susypv} by introducing
matter chiral PV supermultiplets $Z^I,Y_I,\varphi^a$ with gauge
transformation properties:
\beq \del^a Z^I = i(T^a Z)^I,\qquad\del_a Y_I = -i(T^T_a Y)_I,
\qquad\del^a \varphi^b = f^{a b c} \varphi_c,\eeq 
and superpotential
\beq W_{\rm P V} = \half W_{i j}Z^I Z^J + \sqrt{2}g\varphi^a(T_a Z)^iY_I,\eeq
provided the gauge representation of the matter in the SUSY theory satisfies
the constraint
\beq C_M^a = \Tr T_a^2 = \Tr(T^R_a)^2\label{Ccond}\eeq
for some (reducible) real representation $R$ of the gauge group. The
condition \myref{Ccond} is indeed satisfied in the MSSM and its
extensions, as well as in the hidden sectors of all $Z_3$
orbifolds~\cite{joel}, which is the only class of WCHS vacua that has
been thoroughly studied.  In the case of SUGRA with a dilaton
superfield $S$, additional PV chiral superfields as well as PV Abelian
gauge multiplets are needed~\cite{pv} to cancel all the UV
divergences.

The regularized theory would be anomaly free if the PV mass terms
respected the classical symmetries of the SUSY theory; this is not possible
if the theory is anomalous at the quantum level.  The quadratically divergent
part of the one-loop corrected Lagrangian contains a term, generated by chiral
matter loops,
\beq (\L_Q)_\chi\propto \Lambda^2\Tr L_Q = \Lambda^2\[
N\l\Dc\Xa\r - \l\Dc\Tr\Gaa\r - 2(\im s)^{-1}\Tr T^a\],\label{LQ}\eeq
where $\Lambda$ is the UV cut-off, and 
\beq X_\alpha = - {1\over8}\chiproj\Da K, \qquad \Gaa = -
{1\over8}\chiproj\Gamma^i_{i j}\Da Z^j,\eeq
with $\Gamma^i_{j k}$ the affine connection associated with the K\"ahler metric.
PV fields with T-duality invariant masses give no contribution to the 
first two terms in \myref{LQ}, and those with \ux\, invariant masses give
no contribution to the last term. One can restore T-duality (but not \ux\,
invariance) by including
a moduli-dependence in the PV mass terms in the superpotential:
$$\mu\to \mu(T^i) = \prod_i\eta(T^i)^{\omega_i}\mu_0,$$
which could be interpreted as arising from string threshold corrections; 
however these are absent~\cite{noth} in $Z_3$ and $Z_7$ orbifold compactifications.

\subsection{The regulated theory}

In the PV regulated theory, the contribution \myref{LQ} is replaced by
\beq (\L_Q)_\chi\propto \Tr\eta\[L_Q|m(z,\z,\l\vx\r)|^2\],\label{LQ2}\eeq
where $\eta$ is the PV signature, $m$ is the PV mass matrix and
$\vx$ is the \ux\, vector superfield.  The operator \myref{LQ2} is
generally not T-duality and \ux\, invariant; these noninvariant
terms can be canceled ``by hand'', i.e, by imposing conditions on the
signatures and overall coefficients of the masses such that the trace
in \myref{LQ2} vanishes. The cancellation of linear and logarithmic
divergences restricts the PV metric:
$${\mathbf{K}}_{PV}(z,\z,\l\vx\r)$$ 
and therefore the PV masses:
$$ m = e^{K/2}{\mathbf{K}}_{PV}^{-1}\mu.$$
Under T-duality and \ux\, transformations the regulated one-loop Lagrangian
transforms as
\beq \Del \L_{\rm anom} = - \superint\Omega H(T,\Lambda_X)\hc =
{1\over8}\superint{E\over R}\Phi H(T,\Lambda_X)\hc,
\label{DelLa}\eeq
where $\Lambda_X$ is the \ux\, gauge parameter, and
\bea \Omega &=& -\Tr\lbr c_d\[\cM^2\bchiproj\cM^{-2}R^m\hc\] +
{c_g}G_m^{\alpha\dot\beta}G^m_{\alpha\dot\beta} + {c_r}R^m\R^m\rbr
\ddd + {c_w}\Omega_W + \Tr\(c_{a}\Omega^a_{\rm YM} 
- {c_X}\Omega_{X^m}\),\label{Omega}\eea
with
\beq \chiproj\Omega = \Phi, \qquad \bchiproj\Omega = \bar\Phi.\label{Phi}\eeq
The constants $c_i = c_i(\eta,q_n,q_X)$ depend on the
signatures, modular weights $q_n$ and \ux\, charges $q_X$ of the
PV fields, 
$\cM^2$ is a real superfield:
$$ \l\cM^2\r = |m(z,\z,\l\vx\r)|^2,$$
\beq R^m = -{1\over8}\cM^{-2}\chiproj\cM^2, \qquad 
 G^m_{\alpha\dot\beta} = \half\cM[\Da,\Dd]\cM^{-1}
+ G_{\alpha\dot\beta}, \eeq
and the Chern-Simons (CS) superfields $\Omega_i$ are defined by
\bea \chiproj\Omega_W &=& W^{\alpha\beta\gamma}W_{\alpha\beta\gamma},\qquad
\chiproj\Omega_{\rm YM} = \sum_{a\ne X}T^2_a\Wc_a\Wa^a,\nnn
\chiproj\Omega_X^m &=& \Xc_m\Xa^m,\qquad
\Xa^m = {3\over8}\chiproj\Da\ln\cM^2 + \Xa.\label{defOXm}\eea
The CS superfield $\Omega_{X^m}$ can be explicitly constructed~\cite{bgan}
following the procedure~\cite{gg} used for the construction of
$\Omega_{\rm Y M}$.  The chiral superfield strength $W_{\alpha\beta\gamma}$ and
the real superfield $G_{\alpha\dot\beta}$, with $\l
G_{\alpha\dot\beta}\r$ a SUGRA auxiliary field, are related to
elements of the super-Riemann tensor.  The result \myref{Omega} has
been obtained by a component calculation~\cite{bgan} and by a
superconformal superspace calculation, followed by gauge fixing to
\uk\, superspace~\cite{danb}.

\subsection{The (modified) linear supermultiplet}

A linear supermultiplet is defined by the conditions\footnote{See
Section 5 of~\cite{bgg} and references therein.}
\beq (\D^2 - 8\bar R)L = (\bar\D^2 - 8R)L = 0.\label{lin}\eeq
It has three components: the dilaton $\ell = \l L\r$, a fermion, the
dilatino $\chi$, and a two-form $b_{\mu\nu}$ that is dual to the axion
$\im s$; it has no auxiliary field.  The modified
linearity condition replaces \myref{lin} by the conditions 
\beq (\bar\D^2 - 8R)L = -\Phi,\qquad (\D^2 - 8\bar R)L = 
- \overline\Phi,\label{modlin}\eeq
where the chiral superfield $\Phi$ has K\"ahler and Weyl weights
$w_K(\Phi) = 2,\; w_W(\Phi) = 1$, respectively.  Consider a theory
defined by a K\"ahler potential $K$ and a Lagrangian $\L$ of the form
\beq K = k(L) + K(Z,\Z), \qquad  \L = -3\displaystyle{\superint}
\,E\,F(Z,\Z,L).\label{genL}\eeq
A canonical Einstein term for this theory requires
\beq
F- L\frac{\partial F}{\partial L} = 1- \frac{1}{3}L \frac{\partial
k}{\partial L} = - L^2{\pp\over\pp L}\({1\over L}F\),\eeq
which is solved by
\beq {F(Z,\Z,L) = 1 + \frac{1}{3} L V +
\frac{1}{3}L \int \frac{d L}{L} \frac{\partial k(L)}{\partial L}},\label{einsl}\eeq
where $V$ is a constant of integration, independent of $L$.  If we 
take  
\beq V = - bV(Z,\Z) + \del_X V_X\eeq
such that under T-duality and \ux\, transformations
\beq \del V = H + \bar H,\eeq
where $H$ is the holomorphic function introduced in \myref{DelLa},
there is a shift in the tree level Lagrangian \myref{genL}
\beq \Del\L =  {1\over8}\superint{E\over R}\chiproj L H\hc=
-{1\over8}\superint{E\over R}\Phi H\hc = -\del\L_{\rm anom},\eeq
since the first term on the left hand side vanishes under
integration by parts~\cite{bgg}.

\subsection{Chiral/linear duality}

Now consider the Lagrangian
\beq \L_{\rm lin} = -3\superint \, E \[F(Z,\Z,L) 
+ \frac{1}{3} (L + \Omega)(S+\S)\],\label{Llin0}\eeq
where $S = \chiproj\Sigma$ is chiral, with $\Sigma\ne\Sigma^{\dag}$
unconstrained, $L= L^\dag$ is real but otherwise unconstrained, and the
chiral and anti-chiral projections of $\Omega$ are given in
\myref{Phi}.  The equations of motion for $\Sigma,\Sigma^\dag$
\beq {{\pp\L\over\pp\Sigma} = {\pp\L\over\pp\Sigma^\dag}} = 0,\eeq
give the constraints \myref{modlin} on $L$, and \myref{Llin0} reduces to
\bea\L_{\rm lin} &\to& -3\superint\,E F \equiv
\superint\,E\[- 3 + 2L s(L) - L V\], \label{Llin}\\ s(L) &=& - \half
\int \frac{d L}{L} \frac{\partial k(L)}{\partial L},\eea
where the vacuum value $\myvev{s(L)} = g^{-2}$ determines the string scale
coupling constant $g$. Alternatively we can use the equation of motion for $L$:
\beq {\pp\L\over\pp L} = - 3E\lbr{\pp F\over\pp L} + {1\over3}\(S + \S\)
- {1\over3}{\pp k\over\pp L}\[F + {1\over3}L\(S + \S\)\]\rbr = 0,\label{Leom}\eeq
to determine  
\beq L = L(S + \S + V).\label{lofs}\eeq
Once $L$ is eliminated, there are only chiral (and \ux\, vector)
superfields in $F,\;L$ and $K$, and the Einstein normalization
condition on \myref{Llin0} takes the form \myref{lu1k}:
\beq F + {1\over3}L(S + \S) = 1.\label{eins}\eeq
The above duality transformation is valid provided the real superfield 
$\Omega$ has K\"ahler and Weyl weights
\beq  w_K(\Omega) = 0,\qquad w_W(\Omega) = 2,\label{wts}\eeq
so that $E\Omega = E_0\Omega_0$ is Weyl invariant, that is,
independent of $K$ and therefore of $L$.
Combining \myref{eins} with \myref{Leom}, we recover \myref{einsl},
and \myref{Llin0} now becomes
\beq\L_{\rm lin} \to -3\superint\,E\[1 + {1\over3}\Omega(S + \S)\] =
-3\superint\,E + {1\over8}\(\superint{E\over R}S\Phi\hc\)\label{LS}\eeq
 
\subsection{Strategy for Anomaly Cancellation}

A completely regulated supergravity theory was constructed~\cite{bgan}
for the case of three untwisted K\"ahler moduli, which is characteristic
of $Z_3$ orbifold compactification.  The results can be summarized as follows.
\begin{itemize}
\item The K\"ahler potential (wave function) renormalization and
dilaton couplings can be regulated with PV fields with T-duality and
\ux\, invariant masses.
\item The remaining divergences can be regulated by PV fields with 
a simple (T-duality and \ux\, invariant) K\"ahler metric.
\item The generalized modified linearity condition \myref{modlin} is used 
to remove (some or all of) the remaining divergences. The operators in
\myref{Omega} satisfy the requirements \myref{wts}, as can be shown~\cite{danb}
by identifying Weyl invariants in conformal superspace, and then gauge-fixing
to \uk\, superspace.
\item Threshold corrections are incorporated, as appropriate.
\item After performing the duality transformation \myref{Leom}--\myref{LS} to
the chiral formulation for the dilaton, the dilaton K\"ahler potential
takes the form
\beq K(S,\S) = k(S + \S + V), \eeq
which is T-duality and \ux\, invariant since $L$ in \myref{lofs} is invariant, 
implying
\beq \Del S = - H(T,\Lambda_X),\label{delS} \eeq
In this formulation the QFT quantum anomaly is canceled (up to
threshold corrections) by a shift in the tree level Lagrangian
\beq\L_S = - {\superint\(S + \S\)\Omega}\label{newL}\eeq
due to the shift \myref{delS}.
\end{itemize}

The Lagrangian \myref{newL} contains new tree level couplings of the
dilaton.  This is to be expected from superstring theory.  The
two-form potential $b_{\mu\nu}$ of the linear multiplet defined by
\myref{lin} appears through a three-form field strength, its curl:
\beq h_{\mu\nu\rho} = \pp_{[\mu}b_{\nu\rho]}.\eeq
This is modified by \myref{modlin}. The supergravity multiplet of 
10-d SUGRA contains the three-form 
\beq H_{L M N} = \pp_{[L}B_{M N]}  + \omega^{\rm YM}_{M N L}
+ \omega^{\rm Lor}_{M N L},\eeq
which includes the 10-d Yang-Mills and Lorentz Chern-Simons forms.
When the theory is compactified to 4-d SUGRA, the 4-d three-form
includes the 4-d Yang-Mills and Lorentz Chern-Simons forms, as well as
additional terms that arise from contractions of Lorentz indices in
the 6 compact dimensions:
\beq h_{\mu\nu\rho} = \pp_{[\mu}b_{\nu\rho]} + \omega^{\rm YM}_{\mu\nu\rho}
+ \omega^{\rm Lor}_{\mu\nu\rho} + {\rm scalar\; derivatives} + \ldots\eeq
These new couplings can be regulated by PV superfields with invariant mass
terms, as is the case for the dilaton coupling to the gauge sector: $\Phi\to\WaWa$.

Work in progress includes phenomenological applications of the above results,
and tightening their connection to the WCHS.

\section{Acknowledgments}

The results on SUGRA anomalies and their cancellation were obtained in
collaboration with Daniel Butter.  The early stages of this study had
help from Andreas Birkedal, Choonseo Park, Matthijs Ransdorp and
So-Jong Rey, and it has benefited from useful input from Joel Giedt,
Brent Nelson, and Tom Taylor. This work was supported in part by the
Director, Office of Science, Office of High Energy and Nuclear
Physics, Division of High Energy Physics, of the U.S.  Department of
Energy under Contract DE-AC02-05CH11231, in part by the National
Science Foundation under grant PHY-0457315.





\begin{thebibliography}{9}
%
\bibitem{wchs} D.~J.~Gross, J.~A.~Harvey, E.~J.~Martinec and R.~Rohm,
{\em  Nucl.\ Phys.\ B} {\bf 256}, 253--284 (1985).
%
\bibitem{mod}   A.~Giveon, N.~Malkin and E.~Rabinovici,
  {\em Phys.\ Lett.\ B} {\bf 220}, 551--556 (1989);
  E.~Alvarez and M.~A.~R.~Osorio,
  {\em Phys.\ Rev.\ D} {\bf 40}, 1150--1152 (1989).
%
\bibitem{gnrev} M.~K.~Gaillard and B.~D.~Nelson,
{\em  Int.\ J.\ Mod.\ Phys.\ A} {\bf 22} 1451--1588 (2007).
%
\bibitem{bgax}   D.~Butter and M.~K.~Gaillard,
  {\em Phys.\ Lett.\ B} {\bf 612}, 304--310 (2005).
%
\bibitem{gk}   M.~K.~Gaillard and B.~Kain,
  {\em Nucl.\ Phys.\ B} {\bf 734}, 116--137 (2006).
%
\bibitem{rparity}   M.~K.~Gaillard,
  {\em Phys.\ Rev.\ Lett.}  {\bf 94}, 141601 (2005).
%
\bibitem{gs4}     J.~P.~Derendinger, S.~Ferrara, C.~Kounnas and F.~Zwirner,
  {\em Phys.\ Lett.\ B} {\bf 271}, 307--313 (1991);
  G.~Lopes Cardoso and B.~A.~Ovrut,
{\em  Nucl.\ Phys.\ B} {\bf 369}, 351--372 (1992);
M.~Dine, N.~Seiberg and E.~Witten,
  {\em Nucl.\ Phys.}\  B {\bf 289}, 589--598 (1987);
  J.~J.~Atick, L.~J.~Dixon and A.~Sen,
  {\em Nucl.\ Phys.}\  B {\bf 292}, 109--149 (1987).
%
\bibitem{gs}   M.~B.~Green and J.~H.~Schwarz,
  {\em Phys.\ Lett.\ B} {\bf 149}, 117--122 (1984).
 %
\bibitem{th} L.~J.~Dixon, V.~Kaplunovsky and J.~Louis, 
{\em Nucl.\ Phys.\ B} {\bf 355}, 649--688 (1991).   
%
\bibitem{danf}  D.~Z.~Freedman and B.~Kors,
{\em JHEP} {\bf 0611}, 067 (2006);
H.~Elvang, D.~Z.~Freedman and B.~Kors,
{\em JHEP} {\bf 0611}, 068 (2006).
%
\bibitem{pv}  M.~K.~Gaillard,
{\em  Phys.\ Lett.\ B} {\bf 342}, 125--131 (1995);
M.~K.~Gaillard,
  {\em Phys.\ Rev.\ D} {\bf 58}, 105027 (1998);
 M.~K.~Gaillard,
{\em  Phys.\ Rev.\ D} {\bf 61}, 084028 (2000).
%
\bibitem{bgan} D. Butter and M. K. Gaillard, arXive:0906.3503v2
[hep-th], to be pubished in {\em Phys.\ Lett.\ B} 
(DOI:~10.1016/j.physletb.2009.08.023), and paper in preparation.
%
\bibitem{bgw} P.~Binetruy, M.~K.~Gaillard and Y.~Y.~Wu,
{\em Nucl.\ Phys.\ B} {\bf 493}, 27--55 (1997).
%
\bibitem{bd} T.~Banks and M.~Dine,
  {\em Phys.\ Rev.\ D} {\bf 50}, 7454--7466 (1994).
%
\bibitem{fpt}   P.~Fox, A.~Pierce and S.~D.~Thomas,
  ``Probing a QCD string axion with precision cosmological measurements,''
  [arXiv:hep-th/0409059].
%
\bibitem{bgg}   P.~Binetruy, G.~Girardi and R.~Grimm,
  {\em Phys.\ Rept.}  {\bf 343}, 255--462 (2001).
%
\bibitem{susypv}  M.~K.~Gaillard,
  {\em Phys.\ Lett.\ B} {\bf 347}, 284 (1995).
%
\bibitem{joel}   J.~Giedt,
  {\em Annals Phys.}  {\bf 289}, 251--265 (2001).
%
\bibitem{noth}   I.~Antoniadis, K.~S.~Narain and T.~R.~Taylor,
{\em  Phys.\ Lett.\ B} {\bf 267}, 37--45 (1991).
%
\bibitem{gg}  G.~Girardi and R.~Grimm,
{\em  Annals Phys.}\  {\bf 272}, 49--129 (1999).
%
\bibitem{danb} D. Butter, paper in preparation. 
%
\end{thebibliography}

%
\end{document}